# RADIOACTIVITY INDUCED BY NEUTRONS: A THERMODYNAMIC APPROACH TO RADIATIVE CAPTURE

*Alberto De Gregorio*[*]

ABSTRACT – When Enrico Fermi discovered slow neutrons, he accounted for their great efficiency in inducing radioactivity by merely mentioning the well-known scattering cross-section between neutrons and protons. He did not refer to capture cross-section, at that early stage. It is put forward that a thermodynamic approach to neutron-proton radiative capture then widely debated might underlie his early accounts. Fermi had already met with a similar approach, and repeatedly used it.

In 2004, seven decades had elapsed since the artificial radioactivity was discovered. On January 15, 1934 it was announced that the activation of aluminium, boron, and magnesium by α-particles had been obtained in the *Institut du Radium* in Paris. Some weeks later, between the end of February and mid-March, two different laboratories in California showed that deutons and protons as well could induce radioactivity.[1]

On March 25, Enrico Fermi communicated that radioactivity was induced in fluorine and aluminium irradiated with neutrons. In October, a second, crucial discovery was made in the laboratories of the *Regio Istituto Fisico* in Rome: in many cases, neutrons might become more effective if they were slowed down through hydrogenous substances.[2] That especially occurred when a nucleus – particularly a heavy one – became radioactive through a process known as "radiative capture", absorbing a slowed-down neutron and promptly emitting a gamma ray.

It is noteworthy that for some time Fermi, while discussing the effects of slowing down the neutrons, did not refer at all to reaction cross-section between neutrons and nuclei. Rather, he only mentioned scattering cross-section, between neutrons and protons. Only some months later did Fermi theoretically account for radiative capture cross-section to be in inverse proportion to neutron velocity.[3]

In the present paper Fermi's first account of the great efficiency of slow neutrons in inducing radioactivity is reviewed. It is also shown that, though mentioning scattering cross section, his early account can be related to pre-existing evidence of radiative capture and to the early theoretical attempts to explain it. Really, a 'thermodynamic' approach for neutron-proton capture existed, involving detailed balance among neutrons, protons, "diplons", and radiation in equilibrium: it had been widely discussed before October 1934, and its implicit relevance in the early treatment then proposed by Fermi is now put forward. Remarkably, young Fermi had already met with a thermodynamic approach which resembles the one which was later used for neutron radiative capture and deuton photodissociation.

---

[1] I. Curie, and F. Joliot, "Un nouveau type de radioactivité," C. R. Acad. Sci., **198**, 254-56 (1934); C.C. Lauritsen, H.R. Crane, and W.W. Harper, "Artificial production of radioactive substances," Science, **79**, 234-35 (1934); M. Henderson, S. Livingston, and E. Lawrence, "Artificial radioactivity produced by deuton bombardment," Phys. Rev., **45**, 428-29 (1934); H.R. Crane and C.C. Lauritsen, "Radioactivity from carbon and boron oxide bombarded with deutons and the conversion of positrons into radiation," Phys. Rev., **45**, 430-32 (1934); H.R. Crane and C.C. Lauritsen, "Further experiments with artificially produced radioactive substances," Phys. Rev., **45**, 497-98 (1934). It should be pointed out that, in retrospect, the activity that Henderson, Livingston and Lawrence considered as produced by deutons, in some cases was induced by neutrons.

[2] E. Fermi, "Radioattività indotta da bombardamento di neutroni," Ricerca Scientifica, **5** (1), 283 (1934); E. Amaldi, E. Fermi, B. Pontecorvo, F. Rasetti, and E. Segrè, "Azione di sostanze idrogenate sulla radioattività provocata dai neutroni," Ricerca Scientifica, **5** (2), 282-83 (1934).

[3] E. Amaldi, O. D'Agostino, E. Fermi, B. Pontecorvo, F. Rasetti, and E. Segrè, "Artificial radioactivity produced by neutron bombardment – II," Proc. R. Soc. London, **149**, 522-57 (1935).

1. – THE INTERACTIONS BETWEEN NEUTRONS AND PROTONS

An influential conference on nuclear physics was held in Rome in October 1931. George Gamow gave a detailed account of the radiative capture of an alpha-particle by a nucleus (Gamow was not present; Max Delbruck delivered his paper). He estimated its probability to be of the order of only $10^{-5}$ per penetration of the potential barrier. Gamow therefore deemed that a process of nuclear resonance was to be considered, "as in this case the $\alpha$-particle will stay longer inside the nucleus oscillating many times between the walls."[4] But, since resonance levels resulted to be too narrow, Gamow guessed that gamma-emission would more probably take place as a consequence of excitation *without* capture.

Shortly after the Rome Conference, radiative capture and, more in general, secondary gamma-emission by irradiated nuclei became topical again, now involving neutrons as incident particles. Norman Feather, a collaborator of James Chadwick at Cambridge, put forward that important conclusions about the forces acting upon neutrons could be drawn, if the angular distribution of recoiling particles hit by neutrons was investigated in the Wilson chamber. In particular, "hydrogen collisions are so much more suited to this study, owing to the greatly simplified conditions, that this aspect of the problem must be left until the experimental data are available."[5]

According to Chadwick, data about angular distribution of scattered neutrons were still poor. Yet, L.H. Gray and D.E. Lea, two of his collaborators, proved that neutrons scattered by lead showed an approximately isotropic distribution. Moreover, "some preliminary experiments by Mr. Lea, using the pressure [i.e. ionisation] chamber to measure the scattering of neutrons by paraffin wax and liquid hydrogen, suggest that the collision with a proton is more frequent than in other light atoms."[6]

Experiments devoted to neutron scattering rapidly increased in number, revealing, for example, scarce backwards scattering by hydrogenous substances, and an almost isotropic distribution for recoiling protons in the centre of mass system.[7] Chadwick himself came back to those arguments, both in his *Bakerian Lecture*, in May 1933, and in his talk to the Seventh Solvay Conference, in October.[8] He mentioned a formula for the elastic scattering cross-section of neutrons by protons, which according to the wave theory was:

---

[4] George Gamow, "Quantum Theory of Nuclear Structure," in *Convegno di fisica nucleare (ottobre 1931)* (Reale Accademia d'Italia, Rome, 1932), pp. 65-81, on p. 80.

[5] Norman Feather, "The collisions of neutrons with nitrogen nuclei," Proc. R. Soc. London, **136**, 709-27 (1932), on 724.

[6] James Chadwick, "The existence of a neutron," Proc. R. Soc. London, **136**, 692-708 (1932), on 704.

[7] Pierre Auger, and Gabriel Monod-Herzen, "Sur les chocs entre neutrons et protons," C. R. Acad. Sci., **196**, 1102-1104 (1933); John R. Dunning and G.B. Pegram, "Scattering and absorption of neutrons," Phys. Rev., **43**, 497-98 (1933).

[8] James Chadwick, "Bakerian Lecture. The neutron," Proc. R. Soc. London, **142**, 1-25 (1933); James Chadwick, "Diffusion anomale des particules $\alpha$. Transmutation des éléments par des particules $\alpha$. Le neutron," in *Structure et propriétés des noyaux atomiques* – Proceedings of the VII Solvay Conference: Brussels, October 22-29, 1933 – (Gauthier-Villars, Paris, 1934), pp. 81-112.

$$Q = \frac{h^2}{\pi M^2 v^2} \sum_n (2n+1) \cdot \sin^2 \delta_n \tag{1}$$

(where M is the reduced mass, v the relative velocity, and $\delta_n$ are phases depending on M, v, and on the interaction potential V(r)). The uniform angular distribution observed proved that only the spherical harmonic of the zeroth order was important in the formula giving the angular distribution of particles, and therefore only the term with n = 0 survived in Eq. (1); that meant the collision took place in an approximately direct way, and the range of the interaction between neutrons and protons was small with respect to $\lambda = h/2\pi Mv$. If one put $\delta_0 = \pi/2$, one was finally led to the elastic scattering cross-section of neutrons by protons:

$$Q = \frac{h^2}{\pi M^2 v^2} \tag{2}$$

Chadwick noticed that the cross-section showed the expected dependence on neutron velocity; nevertheless, it "est devenue trop grande, puisque … nous obtenons un rayon de choc d'environ $10 \cdot 10^{-13}$ cm, avec une valeur minima de $7 \cdot 10^{-13}$ cm, tandis que les observations donnent plutôt moins que $5 \cdot 10^{-13}$ cm."[9] Chadwick suggested that one could overcome such a discrepancy by invoking a suitable neutron-proton interaction. It is worth noting that neither the deduction of Eq. (2) nor the subsequent arguments invoked by Chadwick persuaded Heisenberg and Fermi, at the Solvay Conference: Heisenberg stressed that, at low velocities, the cross-section should go as the square of the radius of interaction between neutrons and protons, rather than as the square of wavelength of the neutrons. Fermi, on the other hand, objected that an exchange interaction did not lead to the excellent agreement Chadwick had stated; rather, such an interaction "ne suffit pas à améliorer la concordance"[10] between theoretical and experimental results. In any case, Chadwick then remarked that "dans le cas de l'hydrogène, la section est *grosso modo* inversement proportionelle au carré de la vitesse du neutron."[11]

At the end of December, Lea resumed the work he and Gray had begun on neutron scattering in 1932. Results were obtained which, in some respects, resembled those obtained by Rutherford in his classical experiments which had led him to conceive his nuclear model of the atom: by irradiating paraffin wax and liquid hydrogen with neutrons, now Gray and Lea had observed a very intense radiation, emitted even at angles of 120°-180° with the incident direction. Lea commented as follows: "It is clearly impossible for neutrons to be scattered at angles greater than a right angle by single elastic collisions with protons, and calculation shows that multiple scattering cannot explain the observed effects."[12] He used two ionisation chamber, in order to clarify the nature of the backward radiation emitted by hydrogenous specimens: they contained argon and hydrogen respectively, the former being much more sensitive to gamma-rays than the latter. Lea held he could infer the effects of neutrons on hydrogen by merely comparing results obtained with paraffin wax and with graphite. He readily established that heterogeneous electromagnetic radiation, of mean energy between two and four million electronvolts, came from hydrogen irradiated with neutrons. A control experiment showed that the γ-rays from the neutron source should not produce secondary emission, on the score of gamma-rays from thorium C" [$Tl^{208}$] producing no secondary emission.

---

[9] "became too large, since … we obtain a collision radius of about $10 \cdot 10^{-13}$ cm, with a minimum value of $7 \cdot 10^{-13}$ cm, while observations provide a value less than $5 \cdot 10^{-13}$ cm"; Chadwick, "Diffusion anomale" (ref. 8), p. 109.

[10] "does not suffice to improve the agreement"; in *Structure et propriétés* (ref. 8), p. 161.

[11] "in the case of hydrogen, the cross-section approximately goes as the inverse of the square of neutron velocity"; *ibid.*, p. 162. It is interesting to note that, as early as 1929, Ettore Majorana calculated the probability that an alpha particle was captured by a nucleus which had undergone alpha-decay, in such a way that a nucleus preceding in the radioactive genealogy might be reconstituted. He found that the cross-section for the absorption of an alpha-particle with an energy close to a (unstable) nuclear energy level went precisely as the inverse of the square of the alpha-particle velocity. He obtained that result both in the frame of wave mechanics, and in the thermodynamic frame, considering equilibrium between emitted and absorbed alpha-particles. Majorana's analysis was eventually reported in his thesis, Fermi being his supervisor (see S. Esposito *et al.* eds., *Ettore Majorana: Notes on Theoretical Physics* (Kluwer, Dordrecht, 2003), pp. 177-188; a copy of Majorana's (unpublished) thesis is kept at the archives of the University «La Sapienza» of Rome).

> The most plausible way of explaining the results is to suppose that in some of the collisions between the neutron and proton, the particles combine to form $H^2$, the heavy isotope of hydrogen. The combination will result in the emission of energy in the form of gamma-radiation.

Lea found that one out of four interactions between neutrons and protons would end up with the emission of secondary gamma-radiation. He concluded his paper by acknowledging he was indebted to Chadwick for the interpretation of the experimental results.

## 2. – THE CRITICISM TO THE ORIGINAL INTERPRETATION OF LEA'S EXPERIMENTS

Much criticism soon arose, concerning the interpretation of Lea's gamma-radiation as following on the production of deuterium nuclei. Harrie Massey and C. Mohr stated that the probability that neutrons and protons combined together could be calculated assuming that the neutron was a "fundamental charge-free particle", and considering that radiation arose merely from the acceleration of the proton by the field of force of the neutron.[13] Calculations in the dipole approximation for the neutron-proton system depended only slightly on the form assumed for the interaction forces. Massey and Mohr calculated that deuterium nuclei would form once out of one thousand collisions – Lea had observed once out of four –. Moreover, they pointed out that if one considered the neutron rather to consist of one proton and one electron, the dipole moment of the proton-neutron system would vanish, in such a way that the probability of combination, calculated by taking into account exchange interactions, would be further reduced to one event in one million interactions.

On January 22, 1934 a work by Pierre Auger was communicated to the *Académie des Sciences*.[14] The interpretation suggested by Chadwick to Lea was strongly criticised. Firstly, the magnetic analysis showed tracks in a cloud chamber belonging to protons, not to deutons. Also theory was unfavourable. It was hard to understand how a deuton could radiate a gamma-ray with an energy exceeding the binding energy of the deuton itself. Further, the emission probability would be exceedingly small, since the collision would last only $10^{-21}$ seconds. Auger stressed that radiative capture would represent an absolute novelty in nuclear physics: "On ne connaît pas de cas où un noyau capturant une particule émettrait du rayonnement γ sans subir une désintégration."[15]

---

[12] D.E. Lea, "Combination of proton and neutron," Nature, **133**, 24 (1934).

[13] Harrie S.W. Massey, and C.B.O. Mohr, "Radiative collisions of neutrons and protons," Nature, **133**, 211 (1934).

[14] Pierre Auger, "Sur les rayons γ produits par le passage des neutrons à travers les substances hydrogénées," C. R. Acad. Sci., **198**, 365-68 (1934).

[15] "nuclei were never found to capture one particle and emit γ-radiation without disintegrating"; *ibid.*, p. 367.

Gamma-rays from hydrogenous substances showed energies up to some million electronvolts. Still, short tracks in the cloud chamber were much more frequent than long ones, and therefore "il semble bien que les neutrons les plus lents auraient le plus de chances de produire les deutons et le rayonnement émis aurait alors un quantum peu supérieur à 1 MVe."[16] It was well known that the slower the neutrons the more scattered, but radiative capture of less swift neutrons would have resulted in the emission of gamma-rays of too small energies. That led Auger to consider, as Carl Anderson and Irène Curie and Frédéric Joliot had already done,[17] that the proton might consist of a neutron and a positron. In place of the neutron-proton system, the neutron-positron system would be excited and emit radiation. Neutron-positron binding energy was higher than neutron-proton. Hard gamma-ray emission could then be accounted for, and gamma-rays between 2 and 4 MeV eventually conceived. Lea's effect could really be ascribed to fast neutrons, "surtout ceux présentant une résonance avec les niveaux d'excitation des protons."[18]

The case of radiative capture, besides being considered with relation to Lea's results on neutrons and hydrogenous substances, also related to artificial radioactivity investigated at the *Istituto fisico* in Rome.

## 3. – RADIATIVE CAPTURE IN NEUTRON-INDUCED RADIOACTIVITY

On May 10, 1934 Fermi and collaborators proposed that transuranic elements were possibly produced in uranium specimens irradiated with neutrons. The problem of radiative capture therefore began to be dealt with in Rome. It was admitted that activation did not necessarily involve the ejection of material particles, so that the mass number of the irradiated elements might increase by one unit. It was held that "il principio attivo dell'U possa avere numero atomico 93 … ; il processo in questa ipotesi potrebbe consistere in una cattura del neutrone da parte dell'U con una formazione di un $U^{239}$ il quale subirebbe successivamente delle disintegrazioni β."[19] On June 16, Nature published a paper by Fermi, putting forward the possible "capture of the neutron with emission of a quantum, to get rid of the surplus energy."[20] As chemical separation showed, isotopes production following upon neutron irradiation fairly was a general phenomenon, taking place with vanadium, manganese, bromine, iodine, iridium, and gold as well.

In July Fermi went through the difficulties concerning radiative capture,[21] a problem already dealt with in 1931 by Gamow, with relation to α-particles, and in 1934 by Massey and Mohr and by Auger, referring to neutrons. The trouble was that, in the event of radiative capture, the incident particle appeared to remain inside the nucleus for a time too short (apart in case of resonances) to allow photon emission. In the case that isotopes were produced following on neutron irradiation, Fermi's new account then was: the incident neutron might be not captured, the collision rather causing a new neutron to be emitted. Therefore irradiation would end on a radioactive nucleus having the same atomic number of the irradiated nucleus, but its mass number lessened by one unit.

---

[16] "it really seems that the slower the neutrons the higher the probability of producing deutons; emitted radiation will then have a quantum energy only a little higher than 1 Mev"; *loc. cit.*

[17] Carl D. Anderson, "The positive electron," Phys. Rev., **43**, 491-94 (1933); Irène Curie, and Frédéric Joliot, "Électrons positifs de transmutation," C. R. Acad. Sci., **196**, 1885-87 (1933).

[18] "mainly to those showing a resonance with the excitation levels of the protons"; P. Auger, "Sur les rayons γ" (ref. 14), p. 368. It is not completely clear – apart from, possibly, a sort of respect to the Curie and Joliot – the reason why Auger chose the proton instead of the neutron to be described as a complex nucleon.

[19] "the active element of U might have atomic number 93 … ; if this hypothesis holds, then the process might consist in neutron capture by U and the formation of $U^{239}$, which would undergo some β-disintegrations"; E. Amaldi, O. D'Agostino, E. Fermi, F. Rasetti, and E. Segrè, "Radioattività «beta» provocata da bombardamento di neutroni – III," Ricerca Scientifica, **5** (1), 452-53 (1934), on 453.

[20] Enrico Fermi, "Possible production of elements of atomic number higher than 92," Nature, **133**, 898-99 (1934), on 899.

[21] Enrico Fermi, "Radioattività prodotta da bombardamento di neutroni," Nuovo Cimento, **11**, 429-41 (1934).

In 1955 Segrè would recall that a lot of uncertainty wrapped that topic in Rome: "We also found that in many cases a neutron would produce a radioactive isotope of the target and there was great doubt among us whether this would be the result of an (n, 2n) or (n, γ) reaction."[22]

## 4. – DEUTON PHOTODISSOCIATION AND THE THERMODYNAMIC APPROACH TO NEUTRON CAPTURE

Chadwick and Maurice Goldhaber announced deuterium photodissociation in August 1934.[23] They suggested an analogy, between the excitation and the ionisation of atoms on the one hand, and the excitation and the 'ionisation' of complex nuclei by gamma-rays on the other hand. Deuterium, beryllium, and alpha-decaying nuclei were regarded as the best candidates for photodissociation. Really, Chadwick and Goldhaber observed that the 2.6 MeV gamma-rays of thorium C" ($Tl^{208}$) caused deuterium nuclei to break; an ionisation chamber filled with heavy hydrogen revealed the resulting protons. Radium gamma-rays of equal intensity, but of 1.8 MeV energy, could only produce very few kicks instead. Measured cross-section for diplon disintegration by 2.6 MeV gamma-rays led Chadwick and Goldhaber to the approximate value of $10^{-28}$ cm$^2$. They also announced a theoretical work by Hans Bethe and Rudolf Peierls, leading "in the usual quantum mechanical way" to a cross-section in " satisfactory agreement" with their experimental value.

Chadwick and Goldhaber emphasised an interesting link between deuterium photodissociation, and gamma-emission from hydrogenous substances irradiated by neutrons. By considering the radiation observed by Lea as the result of neutrons and protons combining into diplons, Chadwick and Goldhaber described it as the reverse process of the photodissociation they themselves had observed. So they continued:[24]

> Now if we assume detailed balancing of all processes occurring in a thermodynamic equilibrium between diplons, protons, neutrons and radiation, we can calculate, without any special assumption about interaction forces, the relative probabilities of the reaction $[h\nu + {}_1D^2 \rightarrow {}_1H^1 + {}_0n^1]$ and the reverse process.

From the experimental value of the photodissociation cross-section they estimated the radiative capture cross-section for neutrons of 1 MeV, but their calculations gave a capture cross-section much smaller than photodissociation cross-section: "It therefore seems very difficult to explain the observations of Lea as due to the capture of neutrons by protons."

The work by Bethe and Peierls, that Chadwick and Goldhaber cited in August, had been received by the Royal Society on July 26, 1934 and would be published only in 1935.[25] Their approach in some way resembled the one that, in the 1910s, Owen W. Richardson had already adopted for the photoelectric effect, for example in his *Electron Theory of Matter*.[26] We are going through Richardson's thermodynamic approach to the photoelectric effect later, now focusing on Bethe and Peierls' thermodynamic approach to neutron radiative capture.

---

[22] Emilio Segrè, "Fermi and the neutron physics," Rev. Mod. Phys., **27**, 257-63 (1955), on 259. One should bear in mind that the sources emitted both neutrons and gamma-rays, so that the re-emission by the target could not be readily distinguished.

[23] James Chadwick, and Maurice Goldhaber, "A 'nuclear photo-effect': disintegration of the diplon by γ-rays," Nature, **134**, 237-38 (1934).

[24] *Ibid.*, p. 238.

[25] Hans Bethe, and Rudolf Peierls, "Quantum theory of the diplon," Proc. R. Soc. London, **148**, 146-56 (1935).

As we already mentioned, their system comprised protons, neutrons, deutons, and radiation in equilibrium. As a first step, they calculated the theoretical cross-section for the deuterium photodissociation:

$$\sigma \sim 1{,}25 \cdot 10^{-26} (\gamma - 1)^{3/2} \gamma^{-3} \text{ cm}^2 \tag{3}$$

(where $\gamma$ is the ratio of the energy $h\nu$ of the – emitted or absorbed – quantum, to the binding energy $\varepsilon$ of the deuteron). For $\gamma = 2$ the expression (3) had its maximum, that is $\sigma = 1{,}6 \cdot 10^{-27}$ cm$^2$, though Chadwick and Goldhaber's experiments led to $\sigma = 10^{-28}$ cm$^2$.

The next step consisted in calculating the cross-section of the reverse process, of neutron-proton radiative capture. The detailed balance led to the relation $P_{12}/g_2 = P_{21}/g_1$ between the probabilities $P_{12}$ and $P_{21}$ of the direct and of the reverse process respectively, where $g_1$ and $g_2$ represented the statistical weight of the final states. The number of the final states was proportional to $\frac{p^2 dp}{dE} = \frac{Mp}{2}$ for mass particles, and to $\frac{(h\nu/c)^2 d(h\nu/c)}{d(h\nu)} = \frac{(h\nu)^2}{c^3}$ for light quanta (to be multiplied by a factor 2 due to polarisation). The ratio of the neutron capture to the gamma-ray absorption probability was:

$$\frac{4(h\nu)^2}{c^3 Mp} \tag{4}$$

Bethe and Peierls introduced also a factor depending on the inverse of the velocity v of the incident neutron. In fact, "in order to obtain the cross-section we must compare the yield not for equal density but for equal current and so must add a factor c/v." The theoretical capture cross-section $\sigma'$ as a function of the photodissociation cross-section $\sigma$ was:

$$\sigma' = 2 \frac{(h\nu)^2}{c^2 p^2} \sigma = 2 \frac{(h\nu)^2}{Mc^2(h\nu - \varepsilon)} \sigma \tag{5}$$

The theoretical value (3) for the photodissociation cross-section was used in Eq. (5), the capture cross-section thus showing a maximum (still for $h\nu/\varepsilon = 2$) equal to $\sigma' = 2{,}7 \cdot 10^{-29}$ cm$^2$, a very small value: "The capture therefore seems hardly observable."

The calculation reported above might appear nothing but the umpteenth hindrance arising from theory, still lacking in an adequate, quantitative explanation of gamma-emission from hydrogenous substances irradiated with neutrons. However, we shall soon see that radiative capture was being debated among physicists more and more, for example at the London Conference of 1934. Besides, the thermodynamic approach allows us to interpret in an original way Fermi's first explanation for radiative capture he provided when he discovered the effect of hydrogenous substances on neutron-induced radioactivity.

In this paragraph, we finally consider the photodissociation of beryllium, discovered by Leo Szilard and T.A. Chalmers about three weeks after Chadwick and Goldhaber had observed the same effect in deuterium. Szilard and Chalmers irradiated beryllium with gamma-rays from radium; beryllium emitted in turn some radiation, which could induce radioactivity in iodine. They came to the conclusion that gamma-rays – and not only alpha-particles – were able to liberate neutrons from beryllium.[27] They were well aware of the importance of their discovery:[28]

---

[26] Owen W. Richardson, *The Electron Theory of Matter* (University Press, Cambridge, 1914).

> It will be possible to make experiments on induced radioactivity by using the gamma rays of sealed radium containers, which are available in many hospitals … Further, it will be possible to have very much stronger sources of neutrons and to produce thereby larger quantities of radioactive elements by using X-rays from high-voltage electron tubes.

As for their choice of exactly a iodine compound as a target where they searched for induced radioactivity, Szilard and Chalmers laconically mentioned that "for certain reasons, we chose to use as indicators elements which, like iodine, are transmuted in the Fermi effect into their own radioactive isotopes." We shall attempt to go back to the "certain reasons" to which they allude, by considering the proceedings of the London Conference of Nuclear Physics.

5. – THE LONDON CONFERENCE OF 1934

Now that the artificial radioactivity had been discovered and the beta-decay theory published, the Conference held in London from 2 to 6 October, 1934 offered many topics concerning the nucleus to discuss of. Really, Frederick Gowland Hopkins (1935), who in 1930 had succeeded to Rutherford as the President of the Royal Society, in the address speech said:[29]

> It seems but yesterday that the atom was though to be made up of protons and electrons alone; now we hear of the properties of neutrons, positrons, photons, and possibly, I am told, also of neutrinos. More, we have made the acquaintance of artificial radio-active substances and even of atoms new to the universe.

Wide attention was reserved to the results obtained by Lea with paraffin wax. During the conference, Lea's results on hydrogenous substances were also related to the more general topic of gamma emission from substances under neutron irradiation. In tackling that topic, the same 'thermodynamic' approach expounded in Bethe and Peierls' work before the Royal Society was adopted.

The current interpretation of the gamma-rays emission from hydrogenous substances irradiated with neutrons was criticised again by Bethe and Peierls during the Conference:[30]

> The emission of radiation in the recombination of proton and neutron can be calculated from thermodynamic considerations. It is smaller by a factor of the order of 1000 than the photoelectric absorption. The experiments of Lea therefore cannot be explained by this recombination radiation.

Massey, tackling the 'Lea problem' and its thermodynamic approach, pointed out:[31]

> In connection with the communication from Drs Bethe and Peierls I should like to point out that in January of this year Dr Mohr and I calculated the probability of combination of a proton and a neutron to form a diplon. This probability can be related to that of the reverse process discussed in the above communication by purely thermodynamic methods. … It seems that experiment and theory agree for the photoelectric process but, if the experiments of Lea and of Fermi and his collaborators are to be interpreted as due to radiative combinations of neutrons and nuclei, there is a big discrepancy for the reverse process. As this would involve a failure of thermodynamics it would seem that the experiments on the radiation phenomena require some other interpretation.

---

[27] Probably, that was the first time that one resorted to induced radioactivity in order to reveal neutrons, instead of vice versa.

[28] Leo Szilard, and T.A. Chalmers, "Detection of neutrons liberated from beryllium by gamma-rays: a new technique for inducing radioactivity," Nature, **134**, 494-95 (1934), on 495.

[29] Frederick Gowland Hopkins, "Address of welcome," in *International Conference of Physics. Papers & Discussions* – Vol. I, Nuclear Physics – (University Press, Cambridge, 1935), p. 2.

[30] Hans Bethe, and Rudolf Peierls, "Photoelectric disintegration of the diplon," in *International Conference* (ref. 29), pp. 93-94; on p. 93.

[31] *International Conference* (ref. 29), p. 168. Note the reference to the photoelectric effect, with relation to the thermodynamic approach.

It is evident that, concerning radiative capture, Lea's experiments were explicitly associated with Fermi's experiments. Chadwick and Feather, and H.R. Crane and C.C. Lauritsen, also discussed radiative capture, both of neutrons and of charged particles. Their communications all faced the impossibility of the radiative capture to be accounted for, owing to the experimental cross-section being too much bigger than the theoretical cross-section.

Concerning the ways in which nuclides might activate following on neutron absorption, isotopes production showed a certain advantage over heavy charged particles ejection. As for the case of charged particles ejected, the lower was the energy of neutrons, the lower were the energy transmitted and the probability that alphas or protons could be ejected through the coulomb barrier of nuclei. As for isotopes, instead, Szilard pointed out that "one may expect that even slow neutrons will induce radioactivity in elements which, like iodine, transmute in the Fermi effect into their own radioactive isotope."[32]

In his speech Fermi gave a short account of the experiments in progress in Rome on neutron-induced radioactivity, and again mentioned the difficulties which arose when one tried to account for isotopical activation. In fact, a neutron would spend so short a time inside the nucleus, that there should be no room for the emission of the excess energy as a gamma-ray: "We must admit that for some reason which we do not know the neutron can remain in the excited nucleus for a much longer time than that quoted."[33] In some respects, Fermi's statement reflected what Gamow had already said about alpha-particles three years before, at the Rome Conference: for the radiative capture of alpha-particles to occur, Gamow had suggested that the alphas, once captured, would oscillate repeatedly between the walls of the nucleus, and so stay longer inside the nucleus itself – just the time needed for the radiative process to occur –. The difference with Gamow's view was that, now, neutrons could not be bound to the nucleus by the Coulomb barrier.

Fermi himself drew a parallelism between isotopical activation and Lea's experiments: "Completely analogous difficulties of interpretation … present themselves in the explanation of the synthesis of heavy hydrogen observed by Lea."

6. – THE RADIATIVE-CAPTURE CROSS-SECTION

Fermi and collaborators, while investigating artificial radioactivity induced by neutrons, discovered that many activities increased in intensity as a consequence of the slowing down of the neutrons. They explained that phenomenon in the following way:[34]

---

[32] *International Conference* (ref. 29), p. 88. Concerning neutrons from beryllium, it should be remembered photodissociation could not provide so high energy neutrons as (α, n) reactions did.

[33] Enrico Fermi, "Artificial radioactivity produced by neutron bombardment," in *International Conference* (ref. 29), pp. 75-77; on p. 77.

[34] "A possible explanation of these facts seems to be the following: neutrons rapidly lose their energy by repeated collisions with hydrogen nuclei. It is plausible that the neutron-proton cross-section increases as the energy decreases. One might think that after some collisions the neutrons end up by moving in a manner analogous to the molecules which diffuse themselves in a gas, possibly reducing their kinetic energy to that due to thermal excitations. In this way, something similar to a solution of neutrons in water or in paraffin would form around the source. The concentration of such a solution at each point would depend on the intensity of the source, on the geometrical conditions of the diffusion, and on possible neutron-capture processes due to hydrogen or to other nuclei present. It is not ruled out that such a point of view may be important in explaining the effects observed by Lea.

> Una possibile spiegazione di questi fatti sembra essere la seguente: i neutroni per urti multipli contro nuclei di idrogeno perdono rapidamente la propria energia. È plausibile che la sezione d'urto neutrone-protone cresca al calare della energia e può quindi pensarsi che dopo alcuni urti i neutroni vengano a muoversi in modo analogo alle molecole diffondentisi in un gas, eventualmente riducendosi fino ad avere soltanto l'energia cinetica competente alla agitazione termica. Si formerebbe così intorno alla sorgente qualcosa di simile a una soluzione di neutroni nell'acqua o nella paraffina. La concentrazione di questa soluzione in ogni punto dipenderebbe dalla intensità della sorgente, dalle condizioni geometriche della diffusione e da eventuali processi di cattura del neutrone da parte dell'idrogeno o di altri nuclei presenti. Non è escluso che un simile punto di vista possa avere importanza nella spiegazione degli effetti osservati da Lea. Sono in corso indagini su tutto questo complesso di fenomeni.

The passage here reported suggests three remarks. a) When Fermi and collaborators wrote about a cross-section decreasing as the neutron velocity increases, they meant the neutron-proton scattering cross-section, exactly the one which was already known to behave in that way. b) The jargon used in the paper is typical of the kinetic theory of gases and of thermodynamics: multiple scattering, particles which diffuse in a gas, thermal excitations. c) Lea's experiments were explicitly mentioned.

It is plain that the paper makes no reference to the increase of the *capture* cross-section as the velocity decreases: Fermi would suggest an explanation involving capture cross-section only in the subsequent papers, but at the moment he only mentioned *scattering* cross-section. That urges on us the need for a thorough examination of the real meaning of Fermi's early explanation.

Therefore, as a first step let us go back to the thermodynamic approach, and consider the formula (5) which Bethe and Peierls calculated for the radiative-capture cross-section. Put there the experimental value that Chadwick and Goldhaber measured for the photodissociation of deuterium, *i.e.* $\sigma = 10^{-28}$ cm$^2$. Then consider the equilibrium conditions described by Bethe and Peierls, and introduce a process thermalising the motion of the neutrons, in fact a process of the same kind as that proposed by Fermi: the neutrons are subject to a process which reduces their velocity – a very efficient process is meant here, since neutron-proton elastic cross-section becomes increasingly large –. The momentum of the neutrons is then reduced to the value corresponding to thermal excitations, and so the quantity p comparing in Eq. (5) should be put equal to 8.4 keV/c . Therefore p is not largely variable any more, at variance with the hypothesis underlying Bethe and Peierls' approach. Neither does the energy hν of the photons emitted in n-p capture vary anymore, but almost equals the binding energy of the deuton, that is hν = 2.2 MeV. To sum up, by putting these values in Eq. (5) one finds that the radiative-capture cross-section for a neutron by a proton would be $14 \cdot 10^{-24}$ cm$^2$ (the actual value is $0.33 \cdot 10^{-24}$ cm$^2$). That means that, by merely considering thermalisation effects, the cross-section, which was initially too small for three orders of magnitudes, now even becomes too large, for more than one order of magnitude! In any case, one should bear in mind that this new value represents only an upper limit for the capture cross-section. In fact nothing prevents neutrons to be captured before they reach a complete thermalisation, if one rests on the frame of the thermodynamic approach.

Note that, as we have widely considered here, analogies had been repeatedly suggested in 1934 while reviewing both Lea's and Fermi's experiments, in relation to radiative capture.

The occurrence that one was really led to account for the high efficiency of slow neutrons by calling on a 'thermodynamic' approach is confirmed by a proposal by Chadwick and Goldhaber, dating back to the summer of 1934. In fact, in 1972 Goldhaber recalled[35] the events

---

Researches about all these phenomena are in progress" (Amaldi *et. al.*, "Azione di sostanze idrogenate" (ref. 2), 283.

[35] Maurice Goldhaber, "Remarks on the prehistory of the discovery of slow neutrons," Proc. R. Soc. Edimburgh, **70**, 191-95 (1972).

occurred in those months and he pointed out that, in the early manuscript for the paper in which he and Chadwick had reported detailed balancing was leading to too small a capture cross-section, a conjecture was put forward: capture cross-section might become very large, for those neutrons which had their velocity reduced by collisions with hydrogen. The two physicists also conjectured that the carbon in the paraffin wax and the metal in the walls of the ionisation chamber, rather than the hydrogen, would capture the neutrons. Yet, Chadwick and Goldhaber left their speculations out of the final manuscript.

Goldhaber stressed that Fermi did not know what he and Chadwick had guessed about the radiative capture of the less swift neutrons (not slow neutrons, properly speaking). According to Goldhaber, Fermi didn't know about it, either, when he and his collaborators announced the effect of hydrogenous substances on neutron-induced radioactivity. Still, he would not need to know Chadwick and Goldhaber's speculations. As it has been shown here, in fact, if one merely rested on the thermodynamic frame and assumed thermalisation to occur, a very simple calculation allowed to overcome the difficulties related to a too small cross-section.

Edoardo Amaldi stated[36] that Chadwick and Goldhaber's approach could not suffice in inducing Fermi to foresee the effect of hydrogenous substances, since too many processes occurring in the slowing down and in the capture of neutrons were still unknown to him in 1934. However, one should bear in mind that the spirit of the present work is not to gather all the facts it was necessary to know in order to obtain a full account of the properties of slow neutrons. Instead, what is meant here is to point at a possible hint, sufficient to suggest to Fermi a theoretical frame allowing of radiative capture of neutrons. In fact, up to then, the balance of theoretical arguments had rather been against radiative capture.

## 7. – THROUGH RICHARDSON'S APPROACH TO THE PHOTOELECTRIC EFFECT

We are now going through Richardson's thermodynamic approach to the photoelectric effect, especially in view of Richardson's influential part in orienting Fermi's scientific interests.

Fermi was a widely self-taught student, and some of his early papers rose from suggestions from treatises he had gone through. Among his most influential "teachers" – besides for example P. Appel, A. Caraffa, O.D. Chwolson, S.D. Poisson – was Richardson. As Maria Cristina Sassi and Fabio Sebastiani showed, the lines followed in his early studies can be easily recognised in a small notebook, entitled *Alcune teorie fisiche*, reporting physics topics and literature Fermi had been studying.[37] The chapter entitled *Teoria elettronica della materia*, for example, is borrowed from Richardson's *Electron Theory of Matter*,[38] a treatise studied in great detail by the young Fermi. It is easily shown that Richardson's book was a source of inspiration for some of Fermi's first works.[39] In fact, when Richardson treated the electromagnetic mass of a point charge moving with variable velocity, he left out the case of extended distribution of charge since "calculations are very complicated and can only lead to the results obtained previously;"[40] now, the inert mass of a rigid system of charges was exactly the topic Fermi tackled in his first paper. He also mentioned Richardson in another of his early papers, dealing with the electromagnetic mass.[41] Still, the most clear reference is in *Sulla teoria statistica di Richardson dell'effetto fotoelettrico,* where Richardson's thermodynamic approach to the photoelectric effect is pursued and further examined.[42]

---

[36] Edoardo Amaldi, "From the discovery of the neutron to the discovery of nuclear fission," Phys. Rep., **111** (1-4), 1-322 (1984); on 165.

[37] Maria C. Sassi, and Fabio Sebastiani, "La formazione scientifica di Enrico Fermi," Giornale di Fisica, **40** (2), 89-113 (1999). The title of Fermi's notebook reads: *Some physics theories*. Fermi's original manuscript is kept at the Joseph Regenstein Library in Chicago; a copy is kept at the *Archivio Amaldi*, at the University "La Sapienza" of Rome.

[38] Richardson (ref. 26).

[39] See Francesco Cordella, and Fabio Sebastiani, "Il debutto di Enrico Fermi come fisico teorico: i primi lavori sulla relatività (1921-1922/23)," Quaderno di Storia della Fisica, **5**, 69-88 (1999); see also: F. Cordella, A. De Gregorio, and F. Sebastiani, *Enrico Fermi. Gli anni italiani* (Editori Riuniti, Rome, 2001).

[40] Owen W. Richardson, *The Electron Theory of Matter* (University Press, Cambridge, 1916), 2nd ed., p. 255.

In a series of papers dating back to the early 1910s, Richardson had developed a statistical method for dealing with some physical properties of conductors. His fundamental assumption was that electrons were dynamically equivalent to the molecules of a gas. Richardson showed that electron absorption and emission, as well as the Thomson effect, could be treated as reversible phenomena subject to the laws of thermodynamics. His method applied to any kind of particles, irrespective of their electric charge. At small concentration such as that of the atmosphere of electrons surrounding a conductor, electrons even appeared to "exhibit a closer dynamical approximation to the ideal gas than any real gas,"[43] since the correction for their finite size was very much less than that for molecules.

Richardson examined what conclusions the statistical and thermodynamic principles led to when they were applied to photoelectric phenomena. He avoided "discussion of the vexed question of the nature of the interaction between the material parts of the system and the ethereal radiation," and just considered "a statistically steady condition of the ethereal and electronic radiations" inside an insulated system containing real bodies in thermal equilibrium. Then, he investigated how energy divided "between the ethereal vibrations and the material part of the system."[44] He assumed the photoelectric emission to be dependent upon the rate of absorption of the (black body) radiation, and evaluated the rate of electrons returning to the body from the external atmosphere. Therefore, by equating absorption and emission rates for the electrons, he eventually obtained the number of liberated electrons as a function of the radiation frequency. The result was "consistent with the view suggested by Einstein that the light of frequency ν communicates to the electrons an amount of energy hν."[45] Still Richardson, as Planck himself, held that Planck's law did not imply the acceptance of the *Licht-quanten* hypothesis, and therefore his treatment of the photoelectric effect did "not necessarily involve the acceptance of the unitary theory of light."[46]

Still in 1934, concerning the thermodynamic approach, it would be acknowledged "it seems that experiment and theory agree for the photoelectric process."[47] Furthermore, Richardson remarked his method was absolutely general. In fact, "the deduction makes no essential use of the fact that particles have been supposed to be electrically charged; so that similar laws may be expected to characterize the reversible formation of gaseous chemical products under the influence of ætherial radiations."[48]

---

[41] Enrico Fermi, "Sulla dinamica di un sistema rigido di cariche elettriche in moto traslatorio," Nuovo Cimento, **20**, 199-207 (1921); Enrico Fermi, "Correzione di una contraddizione tra la teoria elettro-dinamica e quella relativistica delle masse elettromagnetiche," Nuovo Cimento, **25**, 159-70 (1923).

[42] Enrico Fermi, "Sulla teoria statistica di Richardson dell'effetto fotoelettrico," Nuovo Cimento, **26**, 97-104 (1923).

[43] Owen W. Richardson, "The electron theory of contact electromotive force and thermoelectricity," Phil. Mag., **23**, 263-78 (1912), on 273.

[44] Owen W. Richardson, "Some applications of the electron theory of matter," Phil. Mag., **23**, 594-627 (1912), on 617.

[45] Owen W. Richardson, "The application of statistical principles to photoelectric effects and some allied phenomena," Phys. Rev., **34**, 146-49 (1912), on 147.

[46] Owen W. Richardson, "The theory of photoelectric action," Phil Mag., **24**, 570-74 (1912), on 574.

[47] Massey (ref. 31).

[48] Owen W. Richardson, " The laws of photoelectric action and the unitary theory of light," Science, **36**, 57-58 (1912), on 58.

Richardson also tried to give a comprehensive deduction of the value of the thermionic saturation current from the electronic pressure, and considered that not all the electrons impinging on the metal surface were absorbed. Still, "we do not, as yet, know enough about this phenomenon of reflexion to be able to make proper allowance for it."[49] In any case, he came to the conclusion that "out of a mixed aggregate of incident electrons a greater proportion of the slow ones will be absorbed than of fast ones."[50]

Richardson could only find one critical frequency $\nu_0$. He acknowledged that revealed a "direction in which it is practically certain that the foregoing theory is too much simplified … I hope to be able to return to the discussion of these questions later."[51] In expounding the photoelectric effect, Richardson often left out Planck's formula, since simpler calculations were afforded by Wien's law, which introduced only negligible corrections. He used Wien's law also in *Electron Theory of Matter*. On the other hand, in his paper on Richardson's statistical theory of the photoelectric effect Fermi further pursued the case of Planck's distribution law, which Richardson had generally left out. More to say, Fermi eventually managed to find a whole succession of critical frequencies, multiple of the fundamental one $\nu_0$, which was the only one Richardson had obtained.

## 8. – FERMI'S FURTHER RESORT TO THE THERMODYNAMIC APPROACH

We have just remarked how deep an influence Richardson's treatise had on young Fermi. In this respect, it is interesting to add that Richardson's approach was also echoed in Fermi's later works, after the birth of quantum mechanics. In 1928 Fermi published a book on atomic physics, where he collected the topics of his lectures on theoretical physics.[52] In the ninth chapter, devoted to quantum statistics, he put forward the ionisation equilibrium and the thermionic effect as examples of the entropy constant being applied. Following Saha's treatment of atomic ionisation, he described atomic ionisation as a state of thermal equilibrium between neutral atoms, ions, and the electron gas. As for the thermionic effect, he explicitly mentioned Richardson's theory, and dealt with the concentration of electrons in equilibrium with a hot metal surface, remarking the advantages of the quantum treatment over Richardson's classical approach. He reported the values of the electron concentration in the cavity and of the saturation thermionic current, stressing that the latter counted for more for practical purposes. Still, he left out the 'albedo', that is the fractional number of electrons reflected by the metal surface.

Fermi's attitude to the thermodynamic approach also comes out from his lectures on Theoretical Physics held in Rome during the academic year 1931-32. Vito Camiz, one of his students, took note of Fermi's lectures; moreover, the topics expounded in Fermi's course of lectures are listed on archive records.[53] It comes out that, on May 10, 1932, Fermi expounded the "ionisation thermal equilibrium," and two days later the "thermodynamic theory of the thermionic effect." In those lectures, Fermi followed the same lines of his book on atomic physics of 1928, except that he now took into account electron reflection. Fermi applied the laws of themodynamics in dealing with the thermal ionisation and the thermionic effect again in 1934, in *Molecole e cristalli*; then, again, in his book *Thermodynamics*, where some lectures he had taken at the Columbia University in 1936 were recollected.[54] In the last chapter of his book on thermodynamics, starting again from the entropy constant, Fermi tackled the "thermal ionisation of a gas: the thermionic effect" as examples of application of thermodynamic tools. He left out the albedo, which he had taken into account in *Molecole e cristalli* instead.

---

[49] Richardson (ref. 44), p. 604.

[50] Richardson (ref. 46).

[51] Owen W. Richardson, "The theory of photoelectric and photochemical action," Phil Mag., **27**, 477-88 (1914), on 488.

[52] Enrico Fermi, *Introduzione alla fisica atomica* (Zanichelli, Bologna, 1928).

[53] Vito Camiz was born in Ancona, Italy, on January 1, 1907. He graduated as a civil engineer in 1930. Then he joined Mathematics, and graduated in Mathematics in 1933. During the academic year 1931-32 he attended two courses of lectures by Fermi, one on Theoretical Physics, and the other one on Mathematical Physics. A copy of his notes, recording Fermi's lectures from February 1932, has been kindly donated by his son, Paolo Camiz, to the "Archivio Amaldi" of the Physics Department of the University "La Sapienza" of Rome. As for the lists of the topics tackled in Fermi's lectures, they are recorded on some record books kept in the Archives of the University "La Sapienza" in Rome. I am very grateful to Rossana Nardella, for her kind help in providing Fermi's record books for consultation.

Therefore Richardson's thermodynamic approach, which Fermi had met with in his youth, did not lose its appeal for him even in his maturity.

CONCLUSIONS

In a previous paper,[55] I have already emphasised that the increase of the neutron-proton scattering cross-section, as well as the high efficiency of paraffin wax in slowing down and absorbing neutrons, were well known in 1934. Therefore, Fermi's discovery of the effects due to the slowing down of neutrons was not so unexpected as it has been repeatedly depicted,[56] but grew in the frame of acquired knowledge of the time. I suggested that the increase of the elastic cross-section due to the slowing down of neutrons could lead Fermi – may be unconsciously – to guess an analogous behaviour for the radiative-capture cross-section. As a completion of that previous reconstruction, we have now seen what precise role Fermi himself ascribed to the very scattering cross-section. In fact, Fermi did not mention reaction cross-section, in his early explanation for the experimental evidence that neutrons slowed down by hydrogenous substances increase their efficiency in inducing radioactivity. Rather, he only referred to the well-known neutron-proton elastic cross-section.

At the same time, Fermi's early interpretation can be related to a thermodynamic approach in use in 1934 for neutron radiative-capture cross-section and then widely debated. Also remarkably, young Fermi had already met with a thermodynamic treatment, which was explicitly held to apply to charged as well as to neutral particles. In the following years, he repeatedly resorted to thermodynamic tools – in accounting for the ionisation of alkali vapours and for photoelectricity –. That suggests what strong impact scientific experiences and encounters exercise on such minds as Fermi's during their formative years.

I'm deeply grateful to Fabio Sebastiani, of the University "La Sapienza" of Rome, for his advice in the preparation of the present work. I also thank Salvatore Esposito, of the University "Federico II" of Naples, for his suggestions.

---

[54] Enrico Fermi, *Molecole e cristalli* (Zanichelli, Bologna, 1934); the English translation is *Molecules, crystals, and quantum statistics*, M. Ferro-Luzzi transl., Lloyd Motz ed. (Benjamin, New York – Amsterdam, 1966). Enrico Fermi, *Thermodynamics* (Prentice-Hall, New York, 1937).

[55] Alberto De Gregorio, "Caso e necessità nella scoperta da parte di Fermi delle proprietà dei neutroni lenti," Il Giornale di Fisica, **42** (4), 195-208 (2001); the English version is available on e-print archive: physics/0201028.

[56] Amaldi (ref. 36), 153; Emilio Segrè, *Enrico Fermi. Physicist* (The University of Chicago Press, Chicago and London, 1970), p. 80. See also: Enrico Fermi, *Note e memorie (Collected Papers)*, E. Amaldi *et al.* eds. (Accademia Nazionale dei Lincei – the University of Chicago Press, Roma – Chicago), 2 vols., vol. II (1965), pp. 926-27.